\documentclass[prl,twocolumn,showpacs,amsmath,amssymb,superscriptaddress]{revtex4-1}
\usepackage[english]{babel}
\usepackage{amsmath}
\usepackage{graphicx, nicefrac}
\usepackage{float}
\usepackage{siunitx}
\usepackage [latin1]{inputenc}
\usepackage{dcolumn}% Align table columns on decimal point
\usepackage{bm,graphicx}% bold math
\usepackage{etoolbox}
\usepackage[colorlinks]{hyperref}
\usepackage{url}
\usepackage[normalem]{ulem}
\usepackage[dvipsnames,usenames]{color}
\usepackage[version=3]{mhchem} % Formula subscripts using \ce{}
\usepackage{booktabs}
\usepackage{tabularx}

\begin{document}

\title{Magneto-optical response of magnetic semiconductors EuCd$_2$X$_2$ (X= P, As, Sb)}

\author{S. Nasrallah}
%\email[]{serena.nasrallah@unifr.ch} 
\email[]{serena.nasrallah@tuwien.ac.at}
\affiliation{Department of Physics, University of Fribourg, 1700 Fribourg, Switzerland}
\affiliation{Institute of Solid State Physics, TU Wien, A-1040 Vienna, Austria}

\author{D. Santos-Cottin}
\affiliation{Department of Physics, University of Fribourg, 1700 Fribourg, Switzerland}

\author{F. Le Mardel\'e}
\affiliation{LNCMI; CNRS-UGA-UPS-INSA; 25, avenue des Martyrs, F-38042 Grenoble, France}

\author{I. Mohelsk\'y}
\affiliation{LNCMI; CNRS-UGA-UPS-INSA; 25, avenue des Martyrs, F-38042 Grenoble, France}

\author{J. Wyzula}
\affiliation{Department of Physics, University of Fribourg, 1700 Fribourg, Switzerland}

\author{\\L. Ak\v{s}amovi\'c}
%\email[]{luka.aksamovic@tuwien.ac.at}
\affiliation{Institute of Solid State Physics, TU Wien, A-1040 Vienna, Austria} 

\author{P. Sa\v{c}er}
%\email[]{Psacer.phy@pmf.hr}
\affiliation{Department of Physics, Faculty of Science, University of Zagreb, Bijeni\v{c}ka 32, HR-10000 Zagreb, Croatia}

\author{J.~W.~H. Barrett}
\author{W. Galloway}
\affiliation{Department of Physics, University of Fribourg, 1700 Fribourg, Switzerland}
\affiliation{University of Cambridge, UK}

\author{K. Rigaux}
\author{F. Guo}
\author{M. Puppin}
\affiliation{Lausanne Centre for Ultrafast Science (LACUS), \'{E}cole Polytechnique F\'ed\'erale de Lausanne (EPFL), CH-1015 Lausanne, Switzerland}

\author{\\I. \v{Z}ivkovi\'c}
%\email[]{ivica.zivkovic@epfl.ch}
\affiliation{Institut de Physique, \'{E}cole Polytechnique F\'ed\'erale de Lausanne (EPFL), CH-1015 Lausanne, Switzerland}  

\author{J.H. Dil}
\affiliation{Lausanne Centre for Ultrafast Science (LACUS), \'{E}cole Polytechnique F\'ed\'erale de Lausanne (EPFL), CH-1015 Lausanne, Switzerland}

\author{M. Novak}
%\email[]{mnovak@phy.hr}
\affiliation{Department of Physics, Faculty of Science, University of Zagreb, Bijeni\v{c}ka 32, HR-10000 Zagreb, Croatia}

\author{N. Bari\v{s}i\'c}
%\email[]{Neven.barisic@tuwien.ac.at}
\affiliation{Institute of Solid State Physics, TU Wien, A-1040 Vienna, Austria} 
\affiliation{Department of Physics, Faculty of Science, University of Zagreb, Bijeni\v{c}ka 32, HR-10000 Zagreb, Croatia}  

\author{C.~C. Homes}
%\email[]{homes@bnl.gov}
\affiliation{National Synchrotron Light Source II, Brookhaven National Laboratory, Upton, New York 11973, USA}  

\author{M. Orlita}
%\email[]{Milan.Orlita@lncmi.cnrs.fr}
\affiliation{LNCMI; CNRS-UGA-UPS-INSA; 25, avenue des Martyrs, F-38042 Grenoble, France} 
\affiliation{Institute of Physics, Charles University, CZ-12116 Prague, Czech Republic} 

\author{Ana Akrap}
\email[]{ana.akrap@unifr.ch} 
\affiliation{Department of Physics, University of Fribourg, 1700 Fribourg, Switzerland}
\affiliation{Department of Physics, Faculty of Science, University of Zagreb, Bijeni\v{c}ka 32, HR-10000 Zagreb, Croatia}
\date{\today}

\begin{abstract}

In this study, we identify EuCd$_2$X$_2$ (for X = P, As, Sb) as a series of magnetic semiconductors. We examine how 
the band gap of the series responds to X changing from phosphorus (P), to arsenic (As), and finally antimony (Sb).
We characterize the samples using electronic transport and magnetization measurements.
Based on infrared spectroscopy,  we find that the band gap reduces progressively from 1.23~eV in EuCd$_2$P$_2$, to 0.77~eV in EuCd$_2$As$_2$, and finally 0.52~eV in EuCd$_2$Sb$_2$.
In a magnetic field, all three systems show a strong response and their band gaps decrease at 4~K. This decrease is non-monotonic as we change X. It is strongest in the phosphorous compound and weakest in the antimony compound. For all the three compositions, EuCd$_2$X$_2$ remains a semiconductor up to the highest magnetic field applied (16~T).

\end{abstract}
%  
%\pacs{}
\maketitle

% Introduction 
Recent work in infrared optics and magneto-optics has unveiled that EuCd$_2$As$_2$ is a magnetic semiconductor, with a band gap of 770~meV \cite{DSCotin2023,nelson2024revealing,10.1063/5.0183907,PhysRevB.109.125202,CarmineDFT2023}. 
In a magnetic field, the energy gap is strongly reduced, but the system remains gapped even in high magnetic fields. This finding challenged the previously assumed topological semimetal nature of EuCd$_2$As$_2$, where electronic transport measurements indicated a metal-like behavior at elevated temperatures \cite{PhysRevB.109.125202}, {\em ab initio} calculations indicated the Weyl points \cite{Hua2018,PhysRevB.99.245147,APL2020}, and angle-resolved photoemission spectroscopy (ARPES) measurements revealed a conical shape of the valence band \cite{adv2020}. 

Several recent studies reproduced EuCd$_2$As$_2$ samples with activated resistivity behavior \cite{10.1063/5.0183907, nelson2024revealing,PhysRevB.109.125202}, and confirmed the presence of a significant band gap in EuCd$_2$As$_2$.
The question remains: How does chemically tuning the system's composition affect the ground state?

In the present study, we substituted As with Sb and P to investigate if these sister compounds exhibit similar behavior. We employed infrared and magneto-infrared spectroscopy, complemented with measurements of transport and magnetic properties, to probe the band structure.
Our main finding is that the band gap progresses from 1.23~eV in EuCd$_2$P$_2$, to 0.77~eV in EuCd$_2$As$_2$, and finally 0.52~eV in EuCd$_2$Sb$_2$ at low temperature. In a magnetic field, all three systems exhibit a strong magneto-optical response, leading to a reduction in their respective band gaps. This is most prominent in the phosphorus compound, where the band gap is diminished by 150~meV in a field of 0.8~T. Conversely, the smallest decrease is observed in the antimony compound, where 2~T reduced the band gap by 45~meV. Despite these variations, all three compositions remain semiconducting even under high magnetic fields.

\begin{figure*}[!ht]
 \includegraphics[width=\linewidth]{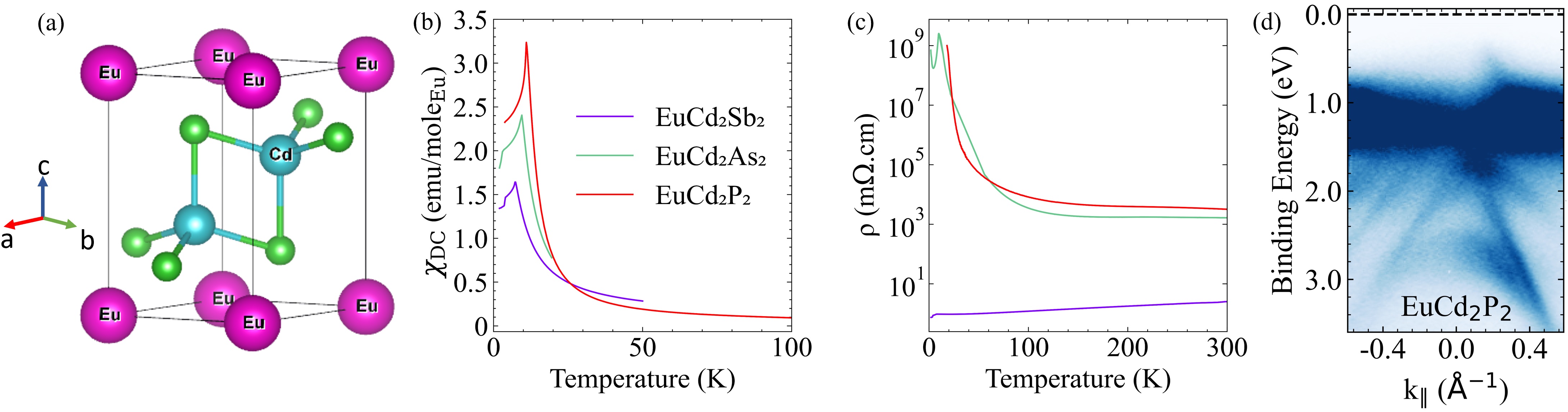}
\caption{ (a) The primitive unit cell of EuCd$_2$X$_2$ (X = Sb, As, P) shows a trigonal crystal structure. 
(b) The magnetic susceptibility while the magnetic field is parallel to the c-axis as a function of temperature measured at 10~Oe for Sb and P and at 100~Oe for As; the pronounced peak indicates the N\'eel temperature (T$_N$) values of 7.3~K, 9.5~K, and 11~K, respectively, for X = Sb, As, and P. 
(c) The resistivity as a function of temperature for EuCd$_2$X$_2$ (X = Sb, As, P). 
(d) Static ARPES band map of EuCd$_2$P$_2$ along the $\overline{M}$-$\overline{\Gamma}$-$\overline{M}$ symmetry point.}
 \label{fig1}
\end{figure*}

%Methods 
Single crystals of high purity EuCd$_2$X$_2$ were synthesized using the Sn flux method \cite{DSCotin2023}. The synthesis process involved a two-step approach, utilizing crystals from the initial growth as seeds.  The trigonal crystal structure of EuCd$_2$X$_2$ is shown in Fig.~\ref{fig1}(a), where the Cd and X layers are situated between the Eu layers \cite{Goryunov2012,chem2011}.
For material characterization, a Physical Properties Measurement System was used to measure resistance up to 300~K. Additionally, characterization included the use of a Superconducting Quantum Interference Device to measure magnetic susceptibility up to 100~K at 10~Oe and 100~Oe, as well as magnetization in a magnetic field up to 7~T at 4~K. 
Angle-resolved photoelectron spectroscopy (ARPES) experiment was carried out at the EPFL LACUS, at the ASTRA end station. The ARPES measurement was carried out at 80~K using a photon energy of 36.8~eV. Single crystals were cleaved \textit{in situ} at a base pressure better than \( 1 \times 10^{-10}~\mathrm{mbar} \). Given the absence of states at the Fermi level in EuCd$_2$P$_2$, polycrystalline Cu was used as a reference for the binding energy.
Infrared reflectance was measured using a Vertex 70v optical spectrometer, while infrared transmission and magneto-transmission measurements were performed using a Vertex 80v optical spectrometer.\\

% results 

%description of Fig.

Peaks in the magnetic susceptibility $\chi_{\mathrm{DC}}$ = M/H (Fig.~\ref{fig1}(b)) at T$_N$ = 11~K \cite{Vsunko2023}, 9.5~K \cite{chem2011}, and 7.3~K \cite{Goryunov2012, HZhang2010} for EuCd$_2$X$_2$ (X = P, As, and Sb), correspond to the antiferromagnetic ordering transition. Around 4~K, a sharp drop in susceptibility indicates the presence of Sn flux on sample surface, as reported in \cite{DSCotin2023}. However, the Sn flux does not affect the sample quality \cite{YueShi2024}. Although the \(\chi_{\mathrm{DC}}\) shows similar behaviour for different X, their electrical resistivity are quite different.
EuCd$_2$Sb$_2$ exhibits a modest metallic behavior at high temperatures \cite{Goryunov2012} with a resistivity of \(2.5 \, \mathrm{m\Omega cm}\) at \(300 \, \mathrm{K}\). We may conclude that the Fermi energy ($E_F$) is located within the valence band. In contrast, EuCd$_2$As$_2$ displays activated behavior with an activation energy of 26~meV and a high resistivity of \(2.5 \times 10^9 \, \mathrm{m\Omega cm}\) at 10~K. 
The resistivity of EuCd$_2$P$_2$ reaches \(9.9 \times 10^8 \, \mathrm{m\Omega cm}\) at 20~K with an activation energy of 16~meV. At lower temperatures the resistivity becomes inaccessible for our measurement. We may conclude that in EuCd$_2$As$_2$ and EuCd$_2$P$_2$ the Fermi energy is within the band gap. The location of $E_F$ is governed by impurity and defect chemistry material synthesis.   

While we were unable to measure the resistive peak in the phosphorus compound, in both arsenide and antimonide the resistive peaks coincide with the antiferromagnetic ordering peaks in the magnetic susceptibility. The increase in resistance near T$_N$ is still not fully understood. One possibility is that above $T_N$, EuCd$_2$As$_2$ and EuCd$_2$Sb$_2$ exhibit enhanced scattering due to magnetic fluctuations \cite{Boothroyd2020}. This effect is more pronounced in As than in Sb, resulting in a stronger peak in As. There is also a suggestion that in EuCd$_2$P$_2$ ferromagnetic clusters are formed above T$_N$ as a result of the spins of charge carriers interacting with the magnetic moments of Eu. These clusters then localize the carriers, leading to an increasing resistance \cite{Vsunko2023, D.YuPRB2024}. To confirm the correct band structure of EuCd$_2$P$_2$, photoemisssion measurement were performed. Figure~\ref{fig1}(d) shows a static ARPES band map of EuCd$_2$P$_2$ at 80~K along the $\overline{M}$-$\overline{\Gamma}$-$\overline{M}$ symmetry point. ARPES data show the presence of Eu $4f$ states as broad features centered around 1-1.2~eV binding energy. At the $\Gamma$ point, there are several hole-like bands, which according to Usachov et al. originate mainly from the p-orbits of phosphorus \cite{D.YuPRB2024}. ARPES data on EuCd$_2$As$_2$ show a similar band map, with $4f$ centered around 1.5~eV \cite{DSCotin2023}.

  \begin{figure*} [!t]
	\includegraphics[width=\linewidth]{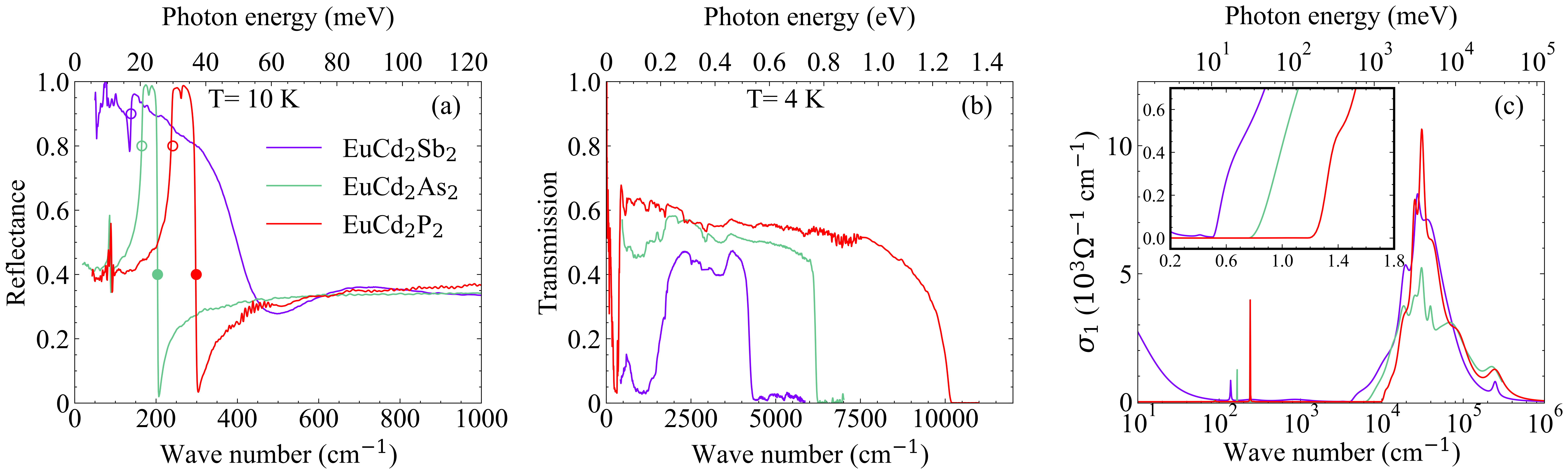}
	\caption{
 (a) Low energy reflectance of EuCd$_2$X$_2$ (X = Sb, As, P) at 10~K . The empty circle indicates the phonon TO modes and the filled one indicates the phonon LO modes. 
 (b) Energy transmission at 4~K exhibits an energy gap of 0.52~eV, 0.77~eV, and 1.23~eV, respectively, for EuCd$_2$X$_2$ (X = Sb, As, P). 
 (c) Multi-layer modeling of the real part of the optical conductivity, $\sigma_1(\omega)$, in the whole energy range. The inset illustrates the optical conductivity in the absorption onset region as a function of wave number.
 }
	\label{fig2}
\end{figure*}

% description of Fig2

All the three compounds belong to the \( P\bar{3}m1 \) space group, so they have four Raman-active phonons and four IR-active phonons within the irreducible vibrational representation, \( \Gamma_{\text{irr}} = 2A_{1g} + 2E_g + 2A_{2u} + 2E_u \) \cite{CCHomes2023}. The \( A_{1g} \) and \( E_g \) modes are Raman active, while the \( A_{2u} \) and \( E_u \) modes are infrared active along the c-axis and within the $ab$ planes, respectively \cite{CCHomes2023}. 
The low-energy infrared reflectance measured at 10~K for light polarized in the $ab$ planes is shown in Fig.~\ref{fig2}(a). The response is different upon changing from Sb to P in the EuCd$_2$X$_2$. In the far-infrared (FIR) range, phonons in EuCd$_2$Sb$_2$ are screened by itinerant carriers, consistent with a metallic resistivity. On the contrary, for EuCd$_2$As$_2$, the reflectance is dominated by two strong in-plane $E_u$ phonon modes at 86 and 165~cm$^{-1}$ \cite{DSCotin2023}. The two phonon modes remain unscreened because free carriers are nearly absent. In the case of EuCd$_2$P$_2$, similar behavior is observed with two $E_u$ phonon modes at approximately 89 and 241~cm$^{-1}$ \cite{CCHomes2023}. Notably, there is a larger longitudinal optical mode and transversal optical mode (LO-TO) splitting in EuCd$_2$P$_2$ than that in EuCd$_2$As$_2$.

In the As and P compounds, the low-frequency phonons are observed at nearby frequencies: 86~cm\(^{-1}\) for As, and 89~cm\(^{-1}\) for P. The lower in energy $E_u$ phonon mode of Sb, highly screened by free charge carriers, was not observed at the level of the signal-to-noise ratio of our data. However, it is expected to occur at frequencies lower than those of As and P, but still in close proximity.
The low-frequency \( E_u \) mode is caused by in-plane atomic displacements that predominantly involve out-of-phase movements between the Eu and Cd atoms, with minimal contribution from the X atoms, which align in phase with the Cd atoms \cite{CCHomes2023}. 
Our data confirms that the X atoms do not significantly affect this oscillation, as indicated by the close frequencies observed in the different compounds.
We next focus on the second \( E_u \) mode that is well observed in all three compounds, with its TO mode indicated by an empty circle and its LO mode by a full circle in Fig.~\ref{fig2}(a). The LO mode is not visible in the Sb compound, because of strong screening. A noticeable change occurs when transitioning from EuCd$_2$Sb$_2$ to EuCd$_2$P$_2$: the energy of the second phonon shifts to a higher value while getting broader. The TO mode of the second phonon for the three compounds is at rather different frequencies, unlike the low energy phonon. This significant difference is attributed to the nature of this second \( E_u \) mode, which mainly involves the out-of-phase displacement of the Sb, As, and P atoms relative to the Cd and Eu atoms. The observed dependence of the phonon energy on the crystal composition correlates well with the atomic masses of Sb, As and P \cite{Tanner_2019}.

More information can be obtained from the reflectivity data, such as $\epsilon_e$, which is the dielectric constant coming from the interband electronic excitations (sometimes referred to as $\epsilon_\infty$). After the second resonance, the reflectivity is approximately flat, with $R_{\infty, Sb} = 0.42$, $R_{\infty, As} = 0.37$, and $R_{\infty, P} = 0.4$. $R_\infty$ is related to $\epsilon_e$, and a flat reflectivity implies that the dielectric function and the refractive index are real \cite{ZrTe5_Akrap}, so we can obtain $\epsilon_e$ from

\begin{equation}
 R = \left|\frac{\sqrt{\epsilon}-1}{\sqrt{\epsilon}+1}\right|^2
\end{equation}

Since all three samples are transparent in the mid-infrared (MIR) and near-infrared (NIR), as shown in Fig.~\ref{fig2}(b), the reflectivity is overestimated in those regions, and therefore $\epsilon_e$ is overestimated. To solve this issue and extract different optical response functions, like the dielectric function or optical conductivity, one can model transmission and reflectance using a multi-layer dielectric response \cite{AKuzmenko2005}. The real part of the optical conductivity, $\sigma_1 (\omega)$ for the whole energy range derived from the model is shown in Fig.~\ref{fig2}(c). The inset of Fig.~\ref{fig2}(c) shows the onset of absorption in $\sigma_1(\omega)$ which coincides with a sudden drop of transmission at $0.52$~eV, $0.77$~eV, and $1.23$~eV for Sb, As, and P, respectively. The sharp onset of absorption due to interband transition indicates a direct band gap \cite{cardona}. From the model, we can get the real part of the dielectric function, $\epsilon_1$ (see \ref{S1} (b)). The flat part in $\epsilon_1$ above the upper $E_u$ phonon mode gives us the correct $\epsilon_e$ of the three compounds: $\epsilon_{e, Sb} = 17.2$, $\epsilon_{e, As} = 11.8$, and $\epsilon_{e, P} = 12.4$. Note that to get $\epsilon_{e}$ of EuCd$_2$Sb$_2$, the Drude contribution had to be removed first.
Another important property we can obtain is the value of static dielectric function $\epsilon_{LST}(0)$ from the Lyddane-Sachs-Teller relation: 

\begin{equation}
\epsilon_{LST}(0) = \epsilon_e \times \left( \frac{\omega_{LO}^2}{\omega_{TO}^2} \right)
\end{equation}

The static dielectric constant is weakly affected by lower energy phonon modes, so we include only the higher energy phonons in its calculation. This gives $\epsilon_{LST, As}(0) = 18$, and $\epsilon_{LST, P}(0) = 18.9$.

\begin{table*}
    \centering
    \renewcommand{\arraystretch}{1.5} % Adjust the vertical spacing
    \begin{tabularx}{\textwidth}{l *{7}{>{\centering\arraybackslash}X}} % Adjusted for horizontal width
        \hline
        \hline
        \textbf{Compound} & \textbf{$T_N$ (K)} & \textbf{$\omega_{TO}$ (cm$^{-1}$)} & \textbf{$\omega_{LO}$ (cm$^{-1}$)} & \textbf{E$_a$ (meV)} & \textbf{E$_g$ (eV)} & \textbf{J$_{\mathrm{eff}}$ (meV)} & \textbf{Volume of primitive cell (\AA$^3$)} \\
        \hline
        EuCd$_2$Sb$_2$ & 7.3 & 138 & - & - & 0.52 & 16 & 148.4 \\
        EuCd$_2$As$_2$ & 9.5 & 165 & 204 & 26 & 0.77 & 82 & 125.0 \\
        EuCd$_2$P$_2$ & 11 & 241 & 298 & 16 & 1.23 & 92 & 114.4 \\
        \hline
    \end{tabularx}
    \caption{The parameters of EuCd$_2$X$_2$ (X = Sb, As, P), where $T_N$ indicates N\'eel temperature, $\omega_{TO}$ and $\omega_{LO}$ indicate the position of phonon TO and LO modes, respectively, $E_a$ represents the activation energy, $E_g$ represents the energy gap, and $J_{\mathrm{eff}}$ is the exchange interaction.}
    \label{tab:my_label}
\end{table*}

\begin{figure*}[!t]
	\centering
	\includegraphics[width=0.75\linewidth]{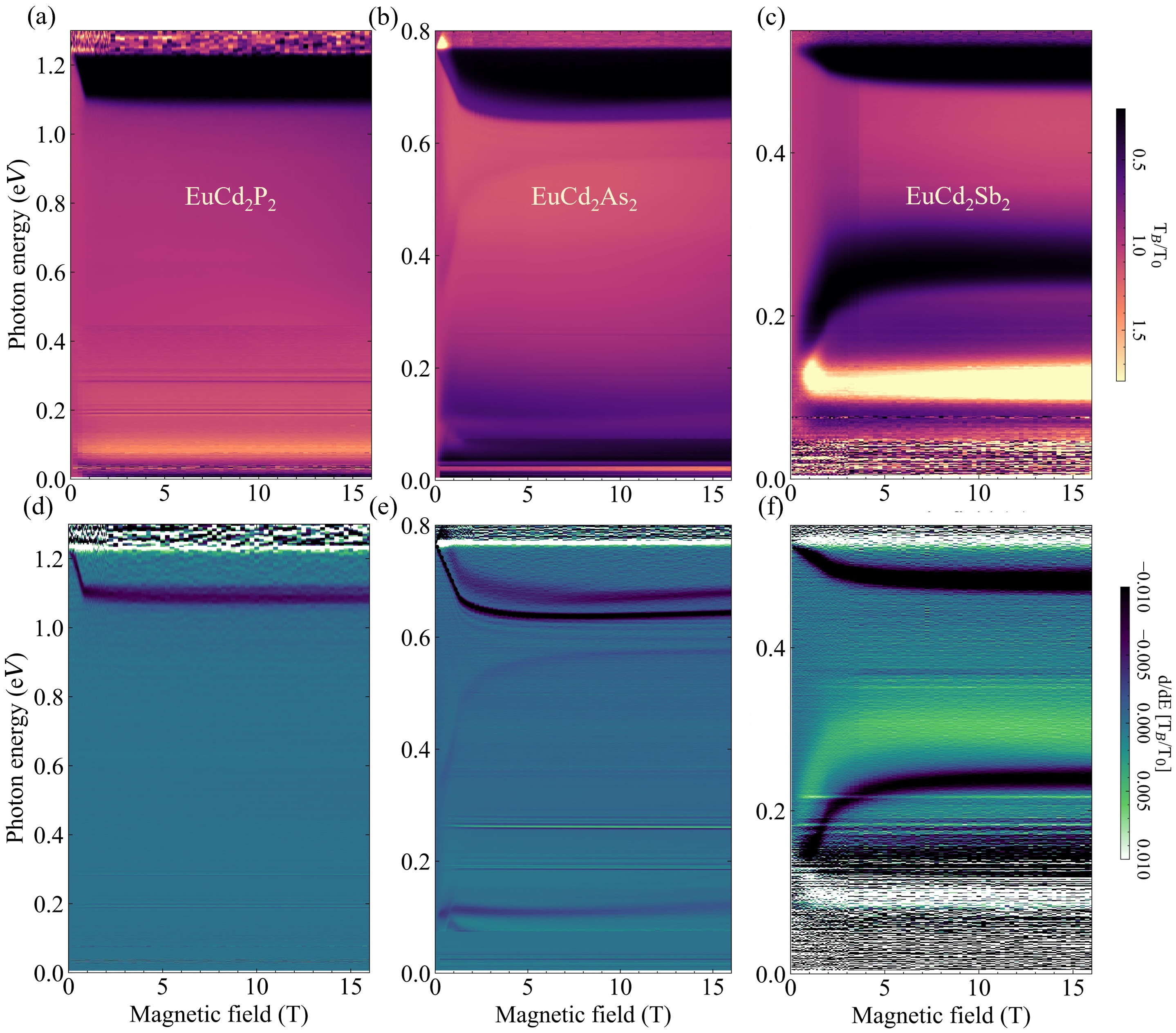}
	\caption{
 Relative magneto-transmission, $T_B$/$T_0$, and the derivative of the relative magneto-transmission $d/dE [T_B/T_0]$ measured at 4.2~K and shown as a function of photon energy and magnetic field for (a-d) EuCd$_2$P$_2$, (b-e) EuCd$_2$As$_2$, and (c-f) EuCd$_2$Sb$_2$. }
	\label{fig3}
\end{figure*}

Applying a magnetic field significantly influences the electronic band structure of EuCd$_2$X$_2$. False color plots in Fig.~\ref{fig3} illustrate these three compounds' relative magneto-transmission across a broad energy and magnetic field range. The energy gap of EuCd$_2$P$_2$ in Fig.~\ref{fig3}(a) undergoes remarkable tuning at a very low magnetic field. At 0~T, the energy gap is approximately 1.23~eV, consistent with the zero-field optics shown in Fig.~\ref{fig2}(a-b). As a magnetic field is applied, the energy gap decreases significantly, reaching 1.08~eV around 0.8~T ($\Delta E_g$ = 150~meV). Beyond 2~T, the reduction of the band gap saturates. Similar behavior is observed in the relative magneto transmission of EuCd$_2$As$_2$ as reported in \cite{DSCotin2023}. The gap decreases by 
130~meV around a field of 1.7~T and reaches a plateau after 5~T. The energy gap in EuCd$_2$Sb$_2$ decreases by 45~meV around a field of 2~T then after 4~T it stays constant. The influence of the magnetic field is also clearly visible in the derivative of the magneto transmission in Fig.~\ref{fig3} (d-f). \textit{Ab initio} calculations performed by Cuono et al. confirmed the energy gap of EuCd$_2$As$_2$ and its evolution in a magnetic field, they based their analysis on the experimental results by Cottin et al. Additionally, they extended these calculations to EuCd$_2$Sb$_2$ and EuCd$_2$P$_2$, obtaining energy gaps that closely match our experimental findings \cite{CarmineDFT2023}.

\begin{figure*}[!t]
	\centering
	\includegraphics[width=0.9\linewidth]{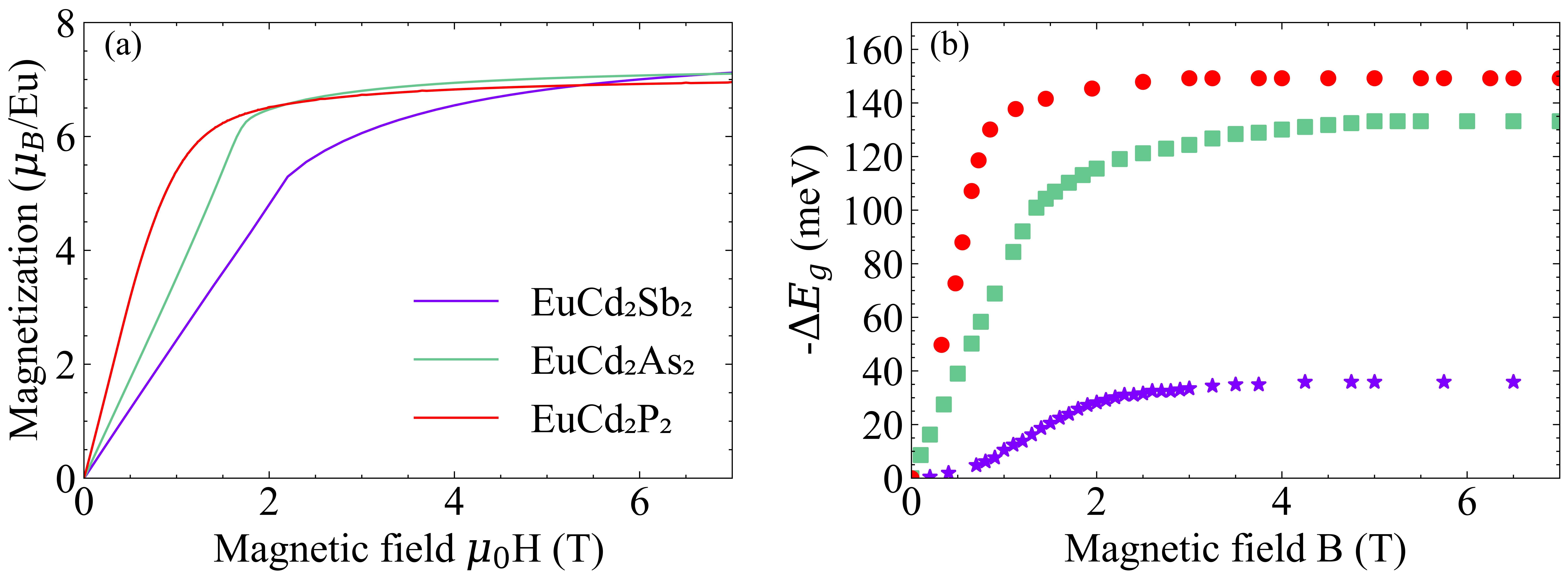} 
	\caption{(a) The magnetization as a function of the magnetic field at 4~K for EuCd$_2$X$_2$ (X = Sb, As, P) with a saturation of around M$_{sat}$ = 7 $\mu_B$. (b) The onset of absorption extracted from the magneto transmission color plots, for  EuCd$_2$X$_2$ (X = Sb, As, P) exhibits a behavior similar to that of the magnetization.}
	\label{fig5}
\end{figure*}

Despite its substantial reduction in the magnetic field, the band gap never closes in any of the three compounds, and no band inversion is observed. Instead, the strong redshift of the band gap, $\Delta E_g(B)$, shown in Fig.~\ref{fig5}(b), seems to be proportional to the magnetization shown in Fig.~\ref{fig5}(a), denoted as $M(\mu_0H)$. This proportionality is observed in europium monochalcogenide materials such as EuTe \cite{Dressel1978}, where there is a connection between the redshift of the absorption edge and the magnetization. The observation of a connection between  $\Delta E_g(B)$ and $M(\mu_0H)$ in our systems validates the molecular field approximation, where $\Delta E_g$ can be expressed as:

\begin{equation}
\Delta E_g = -\frac{1}{2} J_{\text{eff}} \frac{S M(T, H)}{M_S}
\end{equation}

Here, $J_{\text{eff}}$ represents the effective exchange energy between band carriers and Eu spins, $S=7/2$, and $M_S$ is the saturation magnetization. Owing to weak overlap of the $4f$ states with other bands, we expect a minimal hybridization between the Eu bands and bands from other elements. This is why $J_{\text{eff}}$ represents an intra-atomic exchange interaction. Applying this formula to our data, we obtain  $J_{\text{eff}}$ = 92~meV, 82~meV, and 16~meV for EuCd$_2$X$_2$ (X = P, As, and Sb) respectively. This observed exchange interaction could be attributed to an $f-d$ exchange interaction, as discussed in Ref.~\onlinecite{Dressel1978}. The significant change in $J_{\text{eff}}$ from P to Sb indicates that this interaction is notably influenced by the neighboring atoms around Eu. A potential explanation for the decreased $J_{\text{eff}}$ when moving from P to Sb lies in the binding energies of Eu $4f$ states in these compounds or, in other terms, the position of these $4f$ states with respect to the valence band maximum. Specifically, the $4f$ state in P lies at $\sim 1$~eV below the valence band maximum (Fig.~\ref{fig1}(d)), whereas for As it centered around 1.3~eV below the valence band maximum \cite{DSCotin2023}. While ARPES on EuCd$_2$Sb$_2$ has not been reported yet, the {\em ab initio} calculations suggest that the $4f$ states lie even deeper than those of As and P \cite{HaoSu2020_Sb_DFT}. Consequently, the $4f$ states in EuCd$_2$P$_2$ align more closely with the $d$ states in the conduction band than those in EuCd$_2$As$_2$ and EuCd$_2$Sb$_2$, resulting in the observed larger $J_{\text{eff}}$.

The energy gap decreases from phosphorus (P) to antimony (Sb) due to the energy difference between bonding and anti-bonding orbitals, influenced by the overlap of interatomic orbitals. This overlap is affected by the atoms' electronegativity, which increases from Sb to P, as it rises to the right and upwards in the periodic table. Higher electronegativity corresponds to higher bonding energy.
In EuCd$_2$P$_2$, there is a stronger overlap of orbitals compared to EuCd$_2$As$_2$. This stronger overlap leads to a greater energy splitting between bonding and anti-bonding states, indicating higher stability and a higher energy gap for the phosphide compound.

%\section{Conclusion}   
In conclusion, our study investigates the impact of substituting Sb, As, and P on the electrical and optical properties of the material. Replacing nonmetal elements like P with metalloids such as As and Sb induces changes in the band structures, consequently altering the energy gap. Importantly, this substitution does not lead to the closure of the energy gap, and the material retains its magnetic semiconducting nature. 
The observed redshift in the energy gap demonstrates a direct correlation with magnetization. This finding supports the application of the molecular field approximation, revealing that the magnetic properties of Eu significantly influence the band structure of this material family. Therefore,  the magnetic ordering of Eu ions has a direct impact on the band structure of EuCd$_2$X$_2$ (X = Sb, As, P).

%\begin{center}
%\textbf{\large Acknowledgments}
%\end{center}
\section{Acknowledgments}
 This research was supported by the NCCR MARVEL, a National Centre of Competence in Research, funded by the Swiss National Science Foundation (Grant No. 205602). A. A. acknowledges funding from the Swiss National Science Foundation through project No. PP00P2\_202661. N.B. acknowledges the support of the Croatian Science Foundation under Project No. IP-2022-10-3382. M. N. and N. B. acknowledge the support of the CeNIKS project co-financed by the Croatian Government and the EU through the European Regional Development Fund Competitiveness and Cohesion Operational Program (Grant No. KK.01.1.1.02.0013). We acknowledge the support of LNCMI-CNRS, a member of the European Magnetic Field Laboratory (EMFL). The work at the TU Wien was supported by FWF Project P 35945-N. Work at Brookhaven National Laboratory was supported by the Office of Science, U.S. Department of Energy under Contract No. DE-SC0012704.

\section{Appendix}
\begin{figure*} [!t]
    \includegraphics[width=0.9\linewidth]{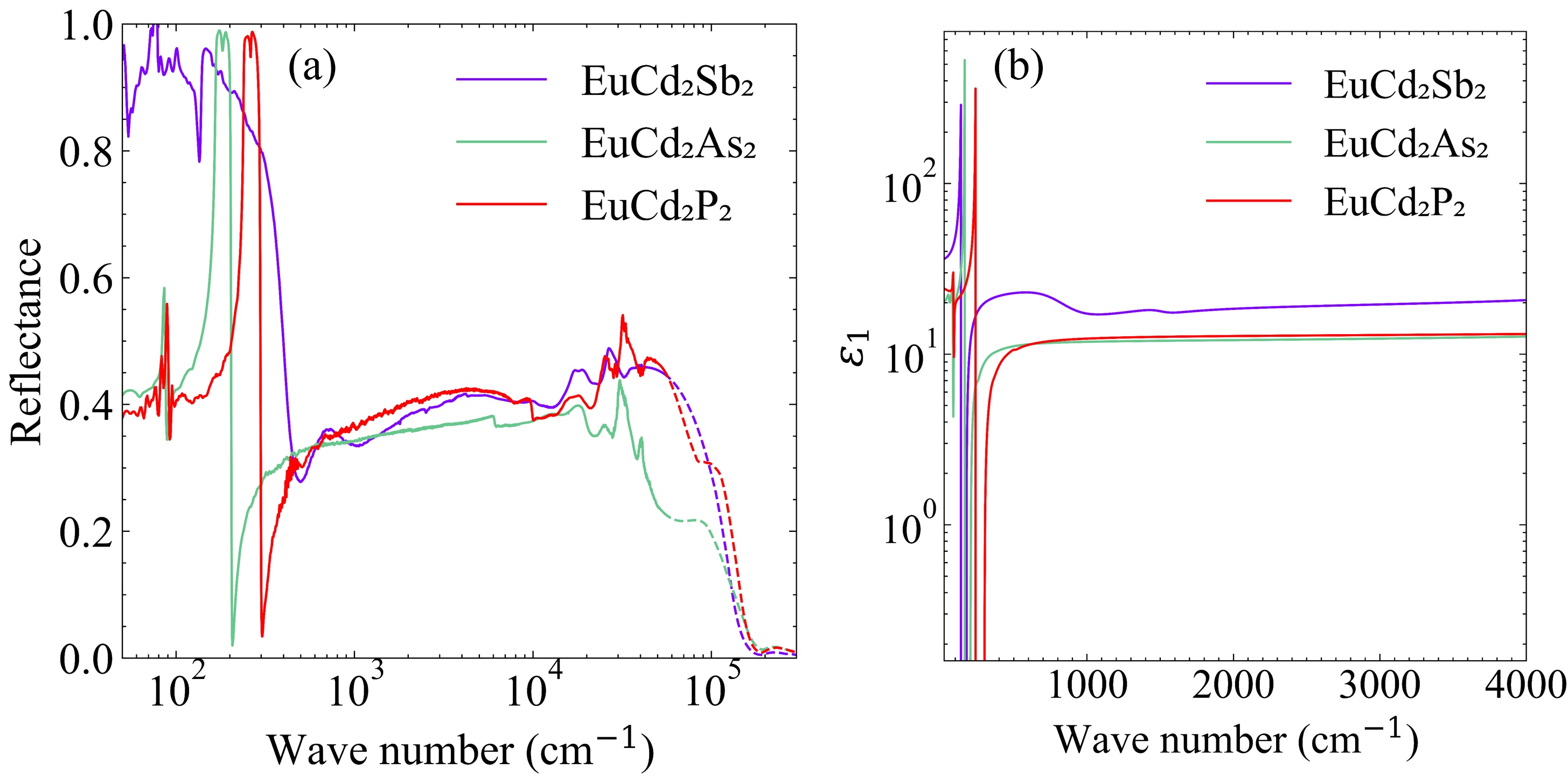}
    \caption{(a) The full range energy reflectance of EuCd$_2$X$_2$ (X = Sb, As, P) at 10~K. (b) The low-energy  real part of the dielectric constant $\epsilon_1$ of EuCd$_2$X$_2$ (X = Sb, As, P) at 10~K extracted from the model.}
    \label{S1}
\end{figure*}

Figure~\ref{S1}(a) shows the reflectance of EuCd$_2$X$_2$ (X = Sb, As, P) over the entire energy range. The low energy extrapolation of the data, below 20 cm$^{-1}$,  is based on an assumption that the reflectance is constant in that regime. The high energy data, above 54000 cm$^{-1}$, is extrapolated using X-ray scattering functions developed by Henke and coworkers \cite{HENKE1993}.

The real part of the dielectric function is plotted in Figure~\ref{S1}(b), where it shows the flat part of $\epsilon_1$ for energies above the second phonon. This dielectric constant was obtained from modeling the reflectivity and transmission.

\begin{figure*} [!t]
    \includegraphics[width=\linewidth]{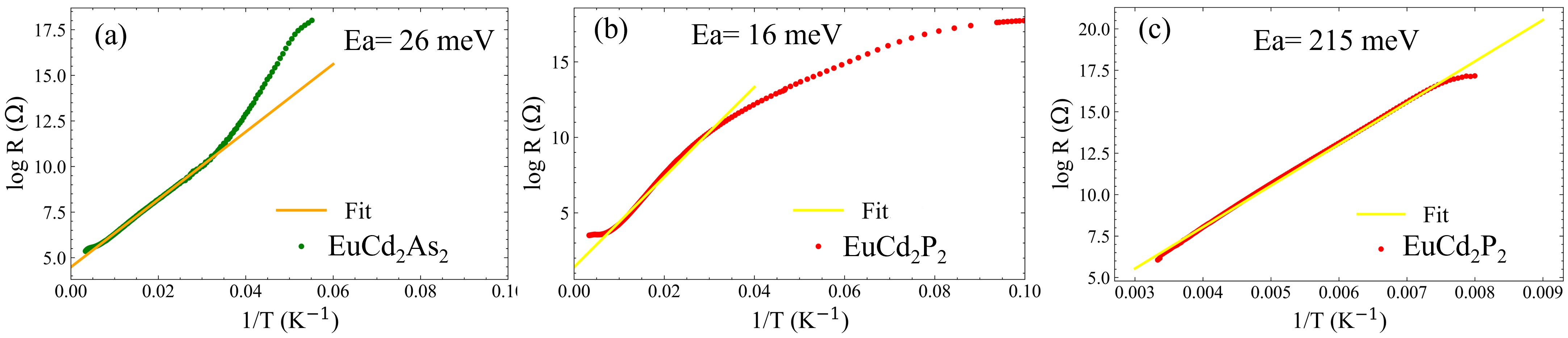}
    \caption{(a) Activation energy extraction by fitting the resistance of EuCd$_2$As$_2$. (b-c) EuCd$_2$P$_2$.}
    \label{S2}
\end{figure*}

Figure~\ref{S2} shows how the activation energy was determined for samples of (a) EuCd$_2$As$_2$ and (b-c) EuCd$_2$P$_2$. We used the Arrhenius equation, $R = R_0 \exp\left(\frac{E_a}{k_B T}\right)$, where $k_B$ is the Boltzmann constant, $E_a$ is the activation energy, and $T$ is the temperature. Samples in Fig.~\ref{S2}(b) and (c) have different carrier concentrations, resulting in very different activation energies. This is probably caused by slighlty different synthesis conditions.

\begin{figure*} [!t]
    \includegraphics[width=0.5\linewidth]{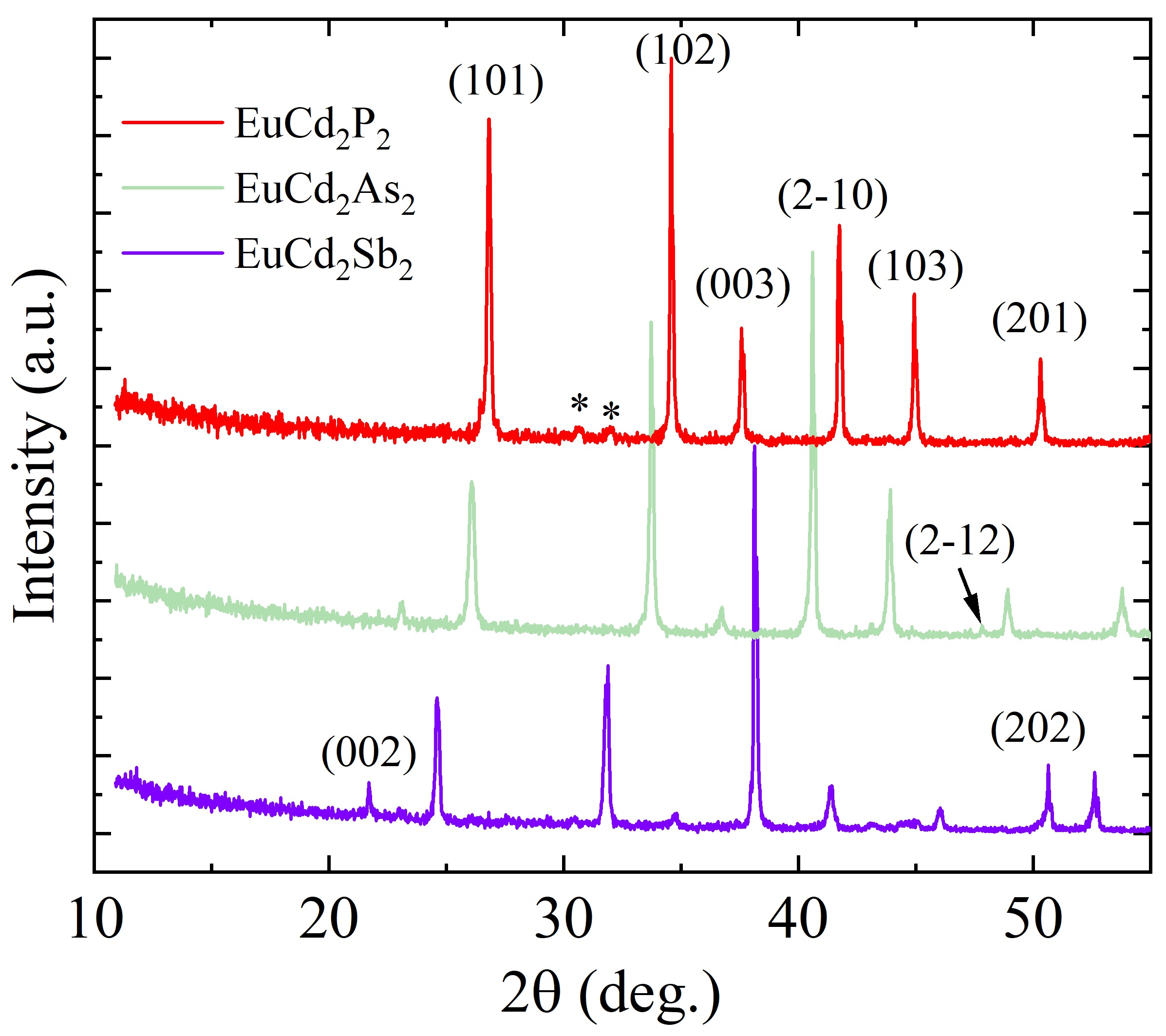}
    \caption{Powder XRD for EuCd$_2$X$_2$ (X=Sb, As, P) obtained from crushed single crystals. All three compounds are isovalent and crystallize in trigonal crystal structure with space group P-3m1 (164). Prominent peaks are marked by Miller indices. Two small peaks  (*) in EuCd$_2$P$_2$ may come from the remaining flux or an impurity phase formed at the crystal boundaries.}
    \label{XRD}
\end{figure*}

\end{document}